# On Nature of Plasmon Drag Effect

M. Durach[1], N. Noginova[2]

[1]Department of Physics, Georgia Southern University, Statesboro, GA 30460
mdurach@georgiasouthern.edu

[2]Center for Materials Research, Norfolk State University, Norfolk, VA 23504
nnoginova@nsu.edu

**Abstract**: Light-matter momentum transfer in plasmonic materials is theoretically discussed in the framework of plasmonic pressure mechanism taking into account non-equilibrium electron dynamics and thermalization process. We show that our approach explains the experimentally observed relationship between the plasmon-related electromotive force and absorption and allows one to correctly predict the magnitude of the plasmon drag emf in flat metal films. We extend our theory to metal films with modulated profiles and show that the simple relationship between plasmonic energy and momentum transfer holds at relatively small amplitudes of height modulation and an approximation of laminar electron drift. Theoretical groundwork is laid for further investigations of shape-controlled plasmon drag in nanostructured metal.

1. Introduction

Two different aspects of light-matter interaction in plasmonic metals are commonly considered in current literature. Firstly, both electromagnetic field and free electrons are many-body entities, which should be viewed in the scope of statistical mechanics. This approach gives rise, in particular, to the emerging field of *hot-electron plasmonics* [1-10], which includes studies of plasmon-induced non-equilibrium electron distributions and thermalization processes in metal [1-7] and interactions of plasmon-generated hot electrons with materials and molecules outside the metal [8-10]. Secondly, quantization of plasmonic oscillators or objects that they interact with has become a hot topic known as *quantum plasmonics*. Several major results in this research area have been obtained in recent years including the prediction and demonstration of coherent stimulated emission of plasmons [11-12], control over spontaneous emission of single quantum emitters by plasmonic nanostructures [13-15], non-classical quantum optics states of plasmons [16-17] and influence of quantum wave properties of metal plasma on plasmons [18-19]. Note that quantum features of these effects are retained in classical or semi-classical considerations.

The photon drag effect is an example of light-matter interaction where the momentum of absorbed light is imparted upon free electrons, and light-induced electric currents are generated. In this respect, the giant enhancement of photon drag effect, known as plasmon drag effect (PLDE), observed in plasmonic films [20-21] and nanostructures [22-26], is of fundamental importance, since it can bring new insights into the aspects of light-matter interaction in metals. From a practical perspective, PLDE opens new avenues for plasmonic-based electronics as it may provide opportunities for incorporation of plasmonic circuits into electronic devices, and for the fields of optical sensors and detectors since it offers a new operational principle and an opportunity to substitute bulky optical detection setups with diffraction limited resolution by compact electronics.

It was conclusively demonstrated that PLDE is closely associated with excitation of surface plasmons [20-26]. In flat silver and gold films, the strongest magnitude of optically induced currents was observed in the conditions of surface plasmon polariton resonance (SPP) [20-22, 24]. Strong photoinduced currents were observed as well in rough metal films and nanostructures at the direct illumination [25-26] with the maximum of the effect close to the localized surface plasmon resonance (LSP) conditions. There were multiple attempts to propose a theoretical mechanism of PLDE associated with SPP excitation. However, compared with experimental results, these predictions were by orders of magnitude larger [20] or smaller [22, 24]. In the case of LSP excitation the two proposed mechanism to-date are the SPIDEr model [23] and the "nano-batteries" model [25], in which the origin of the effect is related to intrinsic nonlinearity of metal in the conditions of LSPs.

In this paper we develop our previously proposed "plasmonic pressure" theoretical mechanism [23], compare theoretical predictions with experiment, and show that PLDE, has both *quantum-optics and hot-electron* aspects, and both of these facets are crucial for the correct description of the effect.

**General Theory of Plasmon Drag Effect**

Classically, light-matter interaction is well described by macroscopic Maxwell equations. The response of matter is represented by the polarization vector $\boldsymbol{P} = \chi \boldsymbol{E}$, where $\chi$ is the susceptibility of the material, the induced polarization charges $\rho = -\nabla \cdot \boldsymbol{P}$ and the currents $\boldsymbol{j} = \partial \boldsymbol{P}/\partial t$. Correspondingly, the Lorentz force density represents the rate of momentum transfer from the field to matter per unit volume as $\boldsymbol{f}_L = -(\nabla \cdot \boldsymbol{P})\boldsymbol{E} + \frac{1}{c}\frac{\partial \boldsymbol{P}}{\partial t} \times \boldsymbol{B}$. We have shown [23] that while the second term, known as Abraham force, is insignificant, the first term can be rewritten as $\boldsymbol{f}_L = \text{grad}(\boldsymbol{P}^c \cdot \boldsymbol{E})$, where superscript "c" exempts a vector from differentiation. This can be represented in components as $f_{L_i} = P_\alpha \partial_i E_\alpha$, where $i, \alpha = x, y, z$ and summation over $\alpha$ is implied. After averaging over an oscillation period $\boldsymbol{f}_L$ has a non-zero rectified component signifying the steady transfer of momentum from SPP fields to electrons with the rate per unit volume obtained as [23]

$$\bar{f}_{L_i} = \frac{1}{2}\text{Re}\{P_\alpha \partial_i E_\alpha^*\}. \tag{1}$$

The effective force given by Eq. (1) can be decomposed as

$$\bar{f}_{L_i} = \frac{1}{2}\text{Re}\chi \cdot \text{Re}\{E_\alpha \partial_i E_\alpha^*\} - \frac{1}{2}\text{Im}\{\chi\}\,\text{Im}\{E_\alpha \partial_i E_\alpha^*\}. \tag{2}$$

The first term corresponds to the striction force, also known as gradient or ponderomotive force. The second term is the electromagnetic pressure force and is solenoidal in nature. Since the first term is curl-free, it produces no overall work on electrons travelling through a circuit, but can result in rectified polarization and serve as a source for intrinsic nonlinearities in metals [27]. Consequently the PLDE electromotive force (emf) driving rectified currents can only stem from the pressure force.

Energy transfer rate (absorption) per unit volume is given by $Q = \frac{\partial \boldsymbol{P}}{\partial t} \cdot \boldsymbol{E}$, and for monochromatic fields after averaging over the period of oscillations,

$$\bar{Q} = -\frac{\omega}{2}\text{Im}\{P_\alpha E_\alpha^*\}. \tag{3}$$

According to Eq. (2), due to high fields and high field gradients in plasmonic conditions, optical forces acting on charge carriers can strongly exceed predictions from the traditional radiation pressure mechanism [23]. They can be very different depending on local positions, but note that for a single-mode field, the time-averaged plasmonic pressure force is related to the time-averaged power. This has a certain similarity to the momentum transfer from photons however with the plasmon wave vector $\boldsymbol{k}_{SPP}$ and plasmon-enhanced absorption. Indeed, comparing the pressure force in Eq. (2) with Eq. (3), one can see that the rectified force density is directly related to the energy transfer as $\bar{\boldsymbol{f}}_L = \hbar \boldsymbol{k}_{SPP} Q_{SPP}/(\hbar \omega_{SPP})$ which can be interpreted from the quantum point of view as the number of quanta absorbed from the plasmonic field multiplied by their momentum, unambiguously signifying the quantum aspect of the problem. The quantum aspect is retained in our classical consideration since the ratio $\hbar k_{SPP}/\hbar \omega$ stays constant in the classical limit $\hbar \to 0$.

## 2. PLDE in flat metal films in Kretschmann geometry

Consider the electric field of a SPP wave propagating in metal films as

$$\boldsymbol{E}(x,z) = \boldsymbol{E}_0(z) e^{i(k_x x - \omega t)}. \tag{4}$$

Following Eq. (2) and taking Eq. (3) into account, the PLDE force density acting in the direction along the film can be calculated as

$$\bar{f}_{L_x}(z) = -\frac{k_x}{2} \text{Im}\{\chi\} |E_0(z)|^2 = \hbar k_x \frac{\bar{Q}}{\hbar \omega}. \tag{5}$$

The momentum transferred from the SPP to electrons can be found as $P_{tr} = \int_{V_{il}, t_p} \bar{f}_{L_x}(z,t) dV dt$ where integration is carried out over the illuminated volume $V_{il}$ of the metal film and the duration of the pulse $t_p$. This momentum is distributed over free electrons with the steady-state value of

$$P_{tr} = n_e V_{il} \frac{t_p}{\tau_m} m^* v_d. \tag{6}$$

Here $n_e$ is the free electron density, $m^*$ is their effective mass, $v_d$ is the resulting drift velocity and $\tau_m$ is the time constant which describes the relaxation of the momentum. We assume that $\tau_m$ is on the order of energy relaxation time, the thermalization time $t_{therm}$. The assumption $\tau_m = t_{therm}$ is confirmed by a very good agreement of the theoretical predictions of Eq. (6) for the PLDE magnitude and its experimental values, as will be shown below. The applicability of the thermalization concept to the relaxation of momentum in excited metal plasma indicates that PLDE is essentially a *hot-electron* plasmonics effect. From this we estimate the PLDE current density $j$ as

$$j = en_e v_d = \frac{e}{m^*} t_{therm} \bar{f}_{L_x}, \tag{7}$$

This result can be also obtained from considering the effect of the plasmonic pressure on the kinetic distribution of electrons, $f = f(v_x, v_\perp)$. For simplicity, here we consider the steady-state conditions of closed circuit (in which the dc polarization field $E_{dcx} \approx 0$) and homogeneous illumination. The distribution function $f$ should depart from the equilibrium distribution $f_0$ and satisfy the steady-state Boltzmann equation

$$\frac{\bar{f}_{L_x}}{n_e m^*}\frac{df}{dv_x} = -\frac{(f-f_0)}{t_{therm}}, \quad (8)$$

Note that the action of the optical forces upon open-circuit metal nanostructure leads purely to induction of dc polarization electric fields, which corresponds to $en_e E_{dcx} = \bar{f}_{L_x}$ as considered in Ref. [23]. In our case the plasmonic pressure induces a strongly non-equilibrium hot-electron distribution which can be found analytically as (see Supplementary Materials for details)

$$f = \int_0^{+\infty} e^{-u} f_0(v_x + uv_p, v_\perp) du = \sum_{n=0}^{\infty} v_d^n \frac{d^n f_0}{dv_x^n} \quad (9)$$

yielding Eq. (7) for the electric current $\boldsymbol{j} = e\int \boldsymbol{v} f(\boldsymbol{v}) d^3 v$. Here $v_d = \bar{f}_{L_x} t_{therm}/(n_e m^*)$. The corresponding PLDE emf, $U$, normalized by the incident intensity $I$, after averaging over the film thickness $h$ can be found as

$$\frac{U}{I} = \frac{j}{\sigma}\frac{L}{I} = \frac{t_{therm}}{\tau}\frac{L/I}{en_e}\frac{1}{h}\int_h \bar{f}_{L_x}(z)dz = \frac{FL}{eI}. \quad (10)$$

Here $L$ is the diameter of the illuminated spot, $F$ is the effective force acting on each electron. We use the experimentally measured conductivity of silver $\sigma = 6.3 \cdot 10^7$ S/m and express our result using the Drude collision time $\tau$. [28]. Using Eqs. (5) and (10) $F$ can be expressed as

$$F = \hbar k_x \frac{t_{therm}}{\tau}\frac{1}{n_e}\frac{dn_{pl}}{dt}, \quad (11)$$

where $\frac{dn_{pl}}{dt} = \frac{1}{h}\int_h \frac{\bar{Q}(z)}{\hbar\omega} dz$ is the rate of plasmonic quanta absorption per unit volume. Below we use $t_{therm} = 1$ ps [1] and $\tau = 31$ fs [28], requiring approximately 30 collisions for electrons to thermalize.

As a numerical example we consider the PLDE observed in a silver film deposited upon a glass prism, as reported in Ref. [21]. The electromagnetic fields are calculated using the analytical solution [29], which predicts the surface plasmon resonance (SPR) at the illumination through the prism at a certain angle of incidence (Kretschmann geometry, [30, 31], Fig 1(a)). The corresponding PLDE pressure force (shown in Fig. 1 (b)) is in the direction of SPP propagation and strongest at the back of the film, reaching $3 \cdot 10^{-19} \frac{\text{N}}{\text{MW/cm}^2}$ per electron.

Let us compare the theoretical predictions of Eq. (10) for PLDE with the experimental data previously obtained in the first PLDE experiments in silver films [21] for two different experimental samples, S1 and S2, see Figure 2 (a) and (b), where the peak of the photoinduced electric signal was observed at SPR conditions. Our theoretical estimations (shown in red) were calculated for the following parameters, corresponding those used in the experiment: film thickness $h = 42$ nm, size of the spot $L = 3$ mm, refractive index of the prism $n_{pr} = 1.78$. One can see a good agreement between theory and experiment in the magnitude of the photoinduced voltage. To our knowledge, this agreement between theory and experiment for PLDE is achieved for the first time.

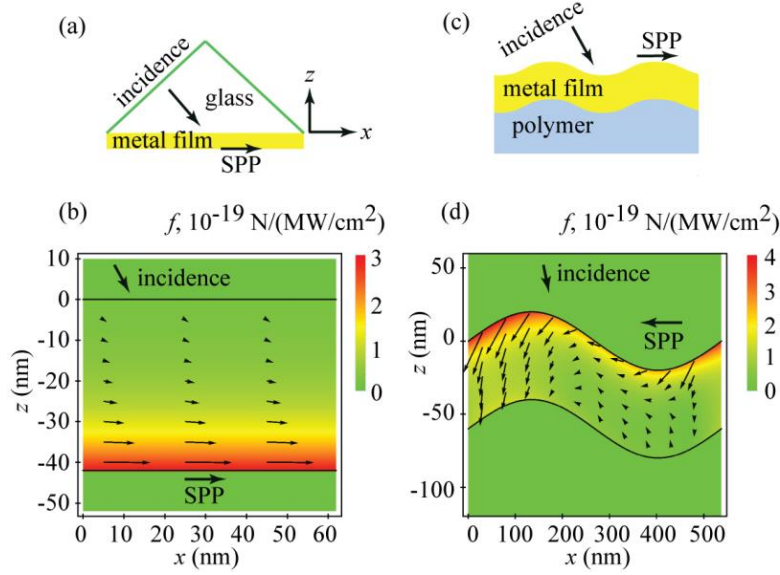

Fig. 1. Plasmonic pressure force in metal films. Flat films: (a) schematic of SPP excitation, (b) plasmonic pressure force distribution (Eq.2) per electron at SPR resonance at $\lambda = 480\ nm$. Films with modulated profile: (c) schematic, (d) plasmonic pressure force distribution at SPR resonance with "backward" propagating plasmon at $\lambda = 608\ nm > d = 538\ nm$.

There are two major differences between the theoretical and experimental data. First, the experimental curves at SPR are significantly broader than simulation results, which can be expected since the calculations assume perfectly flat films. Nevertheless, the angular position of the PLDE peak is reproduced in the theory. Theoretically, the enhanced PLDE emf at SPR is due to enhanced fields that enter Eq. (10) via the force given by Eq. (5), which is proportional to the enhanced absorption rate density $Q$. Second, the off-resonant signal in the experiment has the polarity opposite to the main effect and theoretical predictions, as well as being greater in magnitude than the off-resonance signal in calculations. Nevertheless, this experimental result is completely in line with the general picture of the plasmon drag in the "plasmonic pressure" model. In Figure 2 (c)-(e), we compare the angular profiles of the experimental reflectivity $R(\theta)$ and corresponding profiles of the photoinduced electrical signals $U/I$ for both samples at two different wavelengths. As one can see, in all the cases these profiles practically coincide if a proper offset and scaling are introduced.

The close relationship between $U$ and $A = 1-R$, and the need for an offset $B$ can be explained by taking into account the fact that contributions to both absorbance and PLDE emf come from two kinds of sources: (i) propagating SPPs excited at the resonance conditions with the wave vector $k_x = k_{SPP}$, and (ii) other plasmonic modes excited in the experimental samples due to small scale surface roughness, which include plasmons with high values of $k$.

Consider the ratio,

$$C = \frac{U - U_{rough}}{A - A_{rough}}. \tag{12}$$

Here we offset the full values of $A$ and $U$ by the contributions $A_{rough}$ and $U_{rough}$ from the small scale roughness. In ideally flat films, $U_{rough} = B_{th} = 0$, $A_{rough} = 0$, and PLDE is associated only with SPP excited in Kretschmann geometry. The constant $C$ can be found using Eqs. (10)-(11), $\frac{dn_{pl}}{dt} = \frac{I}{h} A \cdot \cos\theta / \hbar\omega$ and $I = \frac{c}{8\pi} n_{pr} |E_0|^2$, so that

$$C = \frac{t_{therm}}{\tau} \frac{1}{n_e ec} \frac{Ln_{pr}}{h} \sin\theta \cos\theta \approx 6 \frac{mV}{\frac{MW}{cm^2}} \text{ (for our experimental conditions).} \tag{13}$$

In the presence of other plasmon excitations, contributions of plasmonic modes to emf are proportional to their contributions to absorption $U \propto \frac{\hbar k}{\hbar\omega} \cdot A$, according to Eqs. (2)-(3). Our results can be fitted as $U_{spp} \approx U_{rough}$ and $A_{spp} \gg A_{rough}$. This is only possible if $k_{rough} \gg k_{spp}$, which confirms that the offsets come from a small-scale roughness. Note that $A_{rough} \approx 0$ is not quite true for sample 2 as seen in Fig. 2 (d). This fact correlates with the broader SPR in this sample signifying a larger-scale roughness.

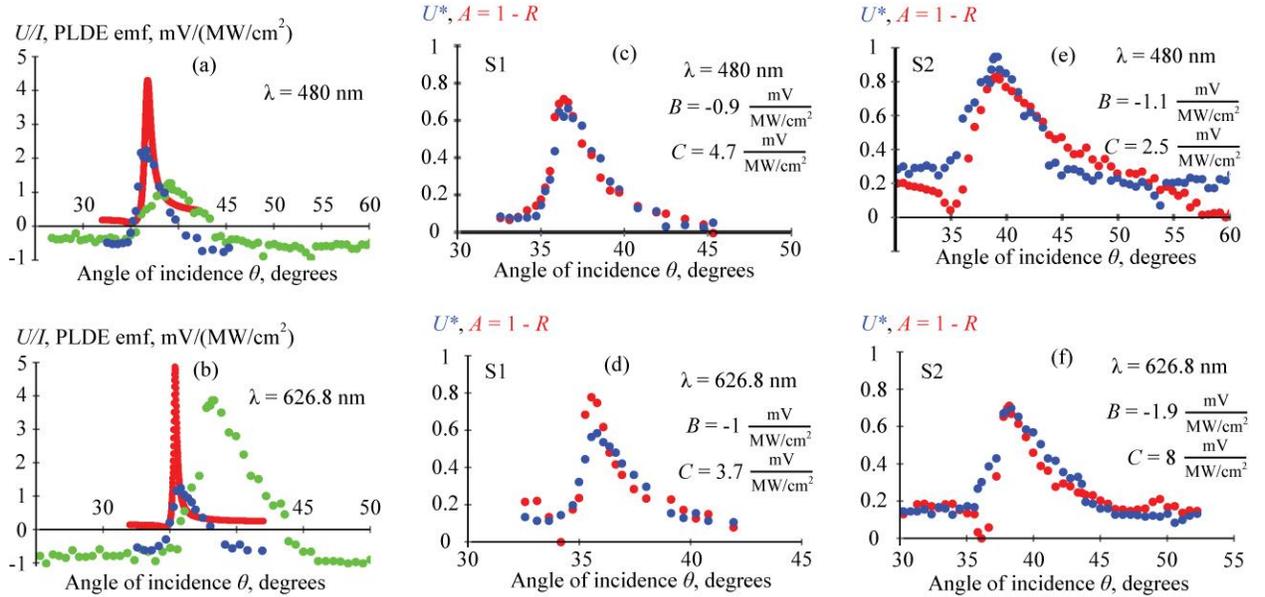

Fig. 2 (a) and (b) PLDE emf in flat silver films, theory (red) and experiment in the samples S1 (blue), S2 (green); (c)-(e) comparison of $U^* = (U/I-B)/C$ (blue, green) and $A=1-R$ (red), the sample, wavelength and parameters B and C are indicated.

One can show that in films with a periodically modulated profile, SPPs propagating in the direction opposite to excitation can be excited in the case of small periods of modulation, $d$, satisfying $\lambda > d > \lambda/(1 + k_{SPP}/k_0)$. The opposite polarity of the off-resonance signal observed in the experiments (negative values of B) can be ascribed to a predominant contribution of such "backward" propagating plasmons generated at the rough surface.

### 3. PLDE in metal films of modulated profile

We would like to extend our model of PLDE to a more complicated geometry including films with surface modulation and investigate if the relationship between PLDE and absorption still holds for multi-mode fields. Consider a metal film with thickness $h$ whose interfaces are given by $z = a(x)$ and $z = a(x) - h$, and $a(x)$ is a periodic function with period $d$. Such structures support SPP excitations, provided that the excitation wave vector satisfies the quasi-momentum conservation $k_x + \frac{2\pi m}{d} = k_{SPP}$, where $m$ is an integer number. Electromagnetic field distribution in such structure can be found using the Chandezon's method [32] which is based on solving Maxwell's equations in a transformed coordinate system with new coordinates $u = x$ and $v = z - a(x)$, in which the field interfaces become flat.

The resulting electric fields in the metal can be represented as [32]

$$E_l = b_j v_{jm}^l e^{ir_j v} e^{i\alpha_m u} = b_j v_{jm}^l e^{ir_j(z-a(x))} e^{i\alpha_m x}. \tag{14}$$

Here index $l = u, v$ characterizes the projection of the field, $v_{jm}^l$ is $m$-th element of $j$-th eigenvector of the Chandezon's method with eigenvalue $r_j$, corresponding to $m$-th diffraction wave with wave number $\alpha_m = k_x + \frac{2\pi m}{d}$. The amplitude of the $j$-th eigenvector, found from the boundary conditions, is $b_j$. In Eq. (14) summation over $j$ and $m$ is implied.

Consider electron drift along a periodically modulated film, characterized by the position-dependent angle $\theta(x, y)$ between electric current and the $x$-axis. Using Eqs. (2), (3) and (14), the work done on electrons by the PLDE pressure force over a period can be found as (see Supplemental Material)

$$\overline{f_{Lx} + \tan\theta \cdot f_{Lz}} \cdot d = \sum_m \frac{\hbar \alpha_m}{\hbar \omega} \overline{Q_m} d + \overline{(\tan\theta - a')f_{Lz}} d - \frac{1}{2}\text{Im}\{\chi\}\overline{a''\text{Im}\{E_x^* E_z\}} d, \tag{15}$$

where bars denote averaging over a period, film thickness, and time and $Q_m$ is absorption of fields in $m$-th diffraction wave. In order to clarify the physical meaning of Eq. (15), let us consider the drift of electrons along trajectories parallel to the film profile $a(x)$. For such electrons the second term on the right-hand side of Eq. (15) vanishes and, if the last term can be neglected (at certain conditions discussed below), the momentum transfer from light to electrons is fully determined by the energy transfer (the first term on the right-hand side of Eq. (15)).

Assume that the direction of electric current in the film is parallel to the film profile $a(x)$, such that $\tan\theta = a'$ and the electrons travel along the film following laminae parallel to each other. With the assumption of such laminar electron current, Eq. (15) allows us to extend Eq. (10) for the PLDE emf on multi-mode plasmonic fields as

$$\frac{U}{I} = \frac{1}{I}\frac{t_{therm}}{\tau}\frac{L}{n_e e}\overline{f_{Lx} + a'f_{Lz}} = \frac{1}{I}\frac{t_{therm}}{\tau}\frac{L}{n_e e}\sum_m \frac{\hbar \alpha_m}{\hbar \omega}\overline{Q_m} = \frac{1}{I}\sum_m U_m. \tag{16}$$

As an example, we consider a sine-wave gold film (Fig.1 (c)) with the profile $a(x) = \Omega \sin(2\pi x/d)$ with the period $d = 538$ nm and amplitude $\Omega = 20$ nm. Substituting fields (14) into Eq. (2) and normalizing per incident intensity we obtain the distribution of the forces inside the metal at SPR conditions, as illustrated at Fig. 1 (d).

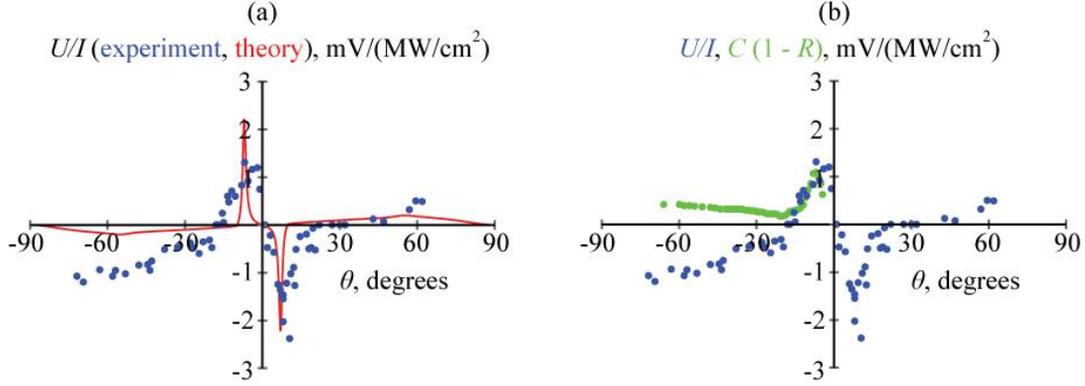

Fig. 3 PLDE emf in the sine-wave gold film (blue circles – two different measurements on the same sample) compared (a) with theory (red trace), (b) with the experimental absorption (green circles) scaled by $C = 3.5 \text{ mV}/(\text{MW}/\text{cm}^2)$.

Our numerical calculations of the PLDE (Eq. (15)) in this structure confirm the direct relationship (16) between the PLDE emf and the energy transfer for both single mode and multimode plasmonic fields [33]. This can be expected in the conditions of predominantly laminar current flow and a relatively small modulation amplitude, $\Omega \ll d$, when derivatives of the profile $a(x)$ can be omitted and the emf is determined primarily by the first term on the right hand side of Eq. (15). However, in a general case, in particular for nanostructured surfaces with a steep height profile, the last two terms in Eq. (15) cannot be excluded and the simple relationship between momentum and energy transfer does not hold. The strict result given by Eq. (15) would allow one to describe or predict PLDE in small-scale, irregular nanostructures, where energy absorption is not directly tied to momentum transfer in the same sense as in this paper, paving a way for additional shape-dependent control and engineering of PLDE.

We compare the calculations to the PLDE measurements performed on sine-wave films in Fig. 3 (a). The experimental data was obtained in a sine-wave gold film with the period of 538 nm and the depth of modulation $2\Omega \sim 50$ nm at the laser light illumination with the wavelength of 630 nm and pulse duration of 5 ns. Details of the structure fabrication, experiment and calculations are presented in [33]. As one can see, similarly to flat films, the theory well describes the magnitude and the angular position of the effect. The experimental dependence is broader than the theoretical predictions, however, there is the direct correlation of the experimental PLDE emf and absorption at SPR in Fig. 3 (b) with the fitting constant $C = 3.5 \text{ mV}/(\text{MW}/\text{cm}^2)$, which coincides with the value for $C$, needed to fit the theoretical PLDE emf (red curve in Fig. 3 (a)) and absorption from the same calculated data, assuming the spot size $L = 2$ mm, and kinetic coefficient for gold, $t_{therm}/\tau = 30$.

### 4. Conclusion

In this paper we demonstrate analytically, numerically and experimentally that rectified drag forces created by plasmonic fields and acting upon electrons in metals, i.e. representing momentum transfer between plasmons and electrons, are intimately related to absorption of these fields, i.e. to respective energy transfer. This relationship follows directly from the *quantum aspect* of energy and momentum transfer between plasmons and electrons, retained in the classical limit. Our theoretical framework and experiments demonstrate that plasmon energy quanta absorbed by the metal plasma are associated with momentum quanta, which are also transferred to electrons upon energy absorption. We demonstrate that in order to correctly predict the magnitude of the experimentally measured PLDE signal, one needs to consider the *momentum relaxation of hot electron*s, which should be on the same time-scale as energy relaxation. By doing so we, to our knowledge for the first time, are able to explain and predict the magnitude of the effect not only qualitatively but in the close quantitative agreement with the experiment. We also show that in a generalized form, this consideration can be extended to complex multi-mode plasmonic field distributions.

**Acknowledgments**

The work was partially supported by PREM grant no. DMR 1205457 and the Army Research Office (ARO) grant W911NF-14-1-0639. M. D. was supported by funds from the Office of the Vice President for Research & Economic Development at Georgia Southern University.